\documentclass[elsart12,epsf,eqsecnum,graphics,cite]{revtex4}
\def\beq{\begin{equation}}

\newcommand{\del}{\partial}

\def\eeq{\end{equation}}
\usepackage{amsmath,amsfonts,amssymb,amsthm,nccmath,latexsym,mathtools}
\usepackage[mathscr]{euscript}
\usepackage{epsfig,epsf}
\usepackage{graphicx}
\usepackage{grffile}
\usepackage{xcolor}

\begin{document}


\voffset1.5cm

\title{On Entanglement Entropy of Maxwell fields in 3+1 dimensions.}
\author{ Candost Akkaya and Alex Kovner}

\affiliation{
Physics Department, University of Connecticut, 2152 Hillside
Road, Storrs, CT 06269-3046, USA}

\begin{abstract}
We consider entanglement entropy between two halves of space separated by a plane, in the theory of free photon in 3+1 dimensions. We show how to separate local gauge invariant quantities that belong to the two spatial regions. We calculate the entanglement entropy by integrating over the degrees of freedom in one half space using an approximation that assumes slow variation of the magnetic fields in longitudinal direction. We find that the entropy is proportional to the transverse area as expected. Interestingly the entanglement properties of the 2D transverse and longitudinal modes of magnetic field  are  quite different. While the transverse fields are entangled mostly in the neighborhood of the separation surface as expected, the longitudinal fields are entangled through an infrared mode which extends to large distances from the entanglement surface. This long range entanglement arises due to necessity to solve the no-monopole constraint condition for magnetic field.
\end{abstract}
\maketitle

\section{Introduction}
In recent years increasing attention is being payed to entanglement aspect of  quantum systems. In particular entanglement entropy between two regions of space in quantum field theories has been a focus of many investigations triggered by the discovery of topological entropy in the context of quantum information theory\cite{Kitaev:2005dm},\cite{PhysRevLett.96.110405}. 

A calculation of von Neumann entanglement entropy is a complicated endeavor \cite{Witten:2018lha} and to date it has been performed either in conformal field theories using CFT methods\cite{Solodukhin:2008dh}, or in free field theories\cite{Casini:2009sr}. Even in free theories this calculation is not entirely straightforward. In particular there is no consensus to date on the result for entropy in  abelian gauge theories\cite{Casini:2013rba},\cite{Casini:2014aia},\cite{Huang:2014pfa},\cite{Huerta:2018xvl}. The early calculation using Euclidean formulation found a nonstandard contact term\cite{Kabat:1995eq} whose existence is still controversial\cite{Donnelly:2015hxa}. In 2+1 dimensions pure Maxwell theory the calculation can be  performed essentially using the equivalence of the theory of a free photon to that of a single massless scalar\cite{Agarwal:2016cir}. In higher dimension however such a tool is not available and it is not a trivial matter how to separate locally physical degrees of freedom in a gauge invariant way\cite{Donnelly:2011hn,Radicevic:2015sza}.

The purpose of this note is to set up the calculation of entanglement entropy in the theory of free Maxwell fields by direct calculation starting from the Schroedinger wave functional of the vacuum. We aim at calculating the reduced density matrix and entanglement entropy in 3+1 dimensional free photon theory by integrating out local gauge invariant degrees of freedom. We consider the simplest situation where we split the space into two halves separated by the $z=0$ plane ("transverse plane"), and integrate local degrees of freedom belonging to one of the halves. Once the correct mode separation is established, we compute the contribution to the entropy due to magnetic field modes with large transverse momentum, or viewed alternatively, modes whose variation in the direction perpendicular to the separation plane is slow. Even in this simple approximation we find nontrivial results. 

We find that the entropy naturally is given by the sum of two contributions - one due to transverse modes of magnetic field and the other due to longitudinal modes. The transverse and longitudinal here are defined in the two dimensional sense relative to the plane separating the two halves of space. This nomenclature arises naturally after solving the condition of vanishing 3D divergence of magnetic field (the no-monopole condition). Both contributions are proportional to the total area of the transverse plane as expected.
The entropy per boundary degree of freedom is finite, but different for transverse and longitudinal modes. 
We also observe that the entanglement properties of the transverse and longitudinal modes are quite distinct.
The transverse modes follow an expected simple pattern - the entanglement between the two halves of space is due to the local modes that live on (or close to) the boundary between the two regions. On the other hand the longitudinal modes exhibit long range entanglement due to a mode which is constant in the direction perpendicular to the boundary.
Technically it arises due to the contribution of the third component of magnetic field $B_3$ which is not an independent degree of freedom, but is determined by
the other two components $B_i,\ i=1,2$ via the solution of the no-monopole condition.


The plan of this paper is the following. In Section II we discuss how to separate local degrees of freedom between the two halves of space in a gauge invariant way. In Section III we perform the calculation of the entanglement entropy in the approximation alluded to earlier. We conclude in Section IV with a short discussion of our results.

\section{Local separation of degrees of freedom}

The object of our study is a theory of a free photon in 3+1 dimensions. We are not going to pay attention to subtleties related to distinction between the compact and noncompact theories\cite{Donnelly:2015hxa} and work directly with the continuum formulation of the theory.

Our first goal is to separate the space into two regions, $\{R:x_3>0\}$ and $\{L:x_3<0\}$ 
where the boundary between the two is the $x_3=0$ plane. In a local field theory this is achieved by separating the full Hilbert space of the theory into a direct product
\begin{align}
\mathcal{H}=\mathcal{H}_L \otimes\mathcal{H}_R
\end{align}
The second step then should be integrating the local degrees of freedom in $R$ in the density matrix describing the vacuum state of the theory (tracing the density matrix over $\mathcal{H}_R$). The reduced density matrix on $\mathcal{H}_L$ is then given by
\beq
\hat\rho_L={\rm Tr}_{\mathcal{H}_R}[\vert\psi\rangle\langle\psi\vert]
\eeq
where $\vert\psi\rangle$ is the vacuum state of the theory.

The splitting into the left and right Hilbert spaces is not straightforward in terms of vector potential degrees of freedom. The
 problem is that one can have for example a  nonvanishing vector potential in the right hand part of space $A^R$ which is a pure gauge and does not lead to nonvanishing local gauge invariant quantities, like the magnetic field $B^R$. Conversely, some configurations with nonvanishing $A^R$ can actually correspond to nonvanishing magnetic fields in the left half space $B^L$, e.g.  magnetic vortices. Thus one cannot simply integrate over $A^R$ if one wishes to separate local gauge invariant degrees of freedom. The integration over all local fields in $R$ must involve integrating over some configurations of vector potential which are nonvanishing in the left half of space. 
 
 This problem can be overcomed if instead of vector potential one works directly with gauge invariant quantities - in the present case the magnetic field. 
Recall that the vacuum wave functional of the theory of a free photon can be written in terms of magnetic field

\begin{align}\label{vacwf}
\langle A|\psi\rangle=\psi_0[\vec{A}]=Nexp\Big\{-\frac{1}{(2\pi)^2}\int d^3xd^3y\frac{B_i(x)B_i(y)}{|x-y|^2}\Big\} \quad\quad i=1,2,3
\end{align}

We will thus strive to decompose the magnetic field of the system as the sum of the field in the regions $R$ and $L$
\begin{align}\label{dec}
\vec{B}(\bm{x})= \vec B^L(\bm{x})+\vec B^R(\bm{x})
\end{align}
with
\beq
\vec B^L(\bm{x})=\vec B(x)\theta(-z); \ \ \ \vec B^L(\bm{x})=\vec B(x)\theta(z)
\eeq

The one subtlety in this strategy is that not all components of the magnetic field are independent, while the functional integral involved in the computation of the trace of the density matrix must be performed over independent degrees of freedom. In particular the components of magnetic field satisfy the 
``no monopole'' condition
\begin{align}
\del_iB_i=0,\ \ \ \ i=1,2,3
\end{align}
This condition must be solved before the functional integral can be properly defined. The geometry of our problem suggests that the simplest way to proceed is to eliminate the longitudinal component of the magnetic field $B_3$ in favor of the transverse components. 
Formally the no monopole condition is solved by
\begin{align}
B_3(x,z)=B_3(x,0)-\int_{0}^z dz' \del_iB_i (x,z')\quad\quad i=1,2
\end{align}
where $x$ stands for the transverse coordinates, and from now on we use $i$ to denote transverse indexes only.
In the following we find it convenient to separate the magnetic field $B_i$ into (two dimensionally) transverse and longitudinal components 
\begin{equation}
\partial_iB_i(x,z)=\chi(x,z); \ \ \ \ \ \epsilon_{ij}\partial_iB_j(x,z)=\zeta(x,z)
\end{equation}

Defining
\begin{equation}
\phi(x)\equiv B_3( x,0); 
\end{equation}
we have (where now by $x$ and $z$ we denote the transverse and longitudinal coordinates respectively)
\begin{align}\label{b3}
B_3(\bm{x},z)=\phi(x)-\int_0^{z} dz'\chi(x,z') 
\end{align}
The integration measure for the functional integral over the magnetic field is thus the Cartesian measure for the (planar) magnetic field $B_i(x,z)$ or equivalently $\chi$ and $\zeta$, and the boundary field $\phi(x)$.

We can now unambiguously split $B_i$ into the field in the right and left half spaces, $B_i(x,z)=B^R_i(x,z)+B^L_i(x,z)$. 
Note that even though $B_3$ is nonlocally related to $\chi$, it is still true that in eq.(\ref{b3}) $B_3^L$ depends only on $\chi_L$ while $B_3^R$ only on $\chi_R$. Thus integrating over $\chi_R$ and $\zeta_R$ is indeed equivalent to integrating over all local fields in the right half space.

As for the boundary field $\phi(x)$ it is up to us which region of space to assign it to. If we assign it to $R$ and integrate over it when calculating the reduced density matrix, this corresponds indeed to integrating over all local gauge invariant degrees of freedom in the right half space, including all possible configurations of the longitudinal field $B^R_3$. As a result however one also integrates over the "source" for the field $B^L_3$  in the left half space. 
Alternatively one can assign $\phi$ to the left half space. This corresponds to not integrating over all possible values of the magnetic field $B^R_3$, but instead allowing all possible configurations of $B^L_3$ to fluctuate in an unrestricted way in the reduced density matrix. It is simply a matter of choice which degrees of freedom to integrate out, and the reduced density matrix as well as entropy will depend somewhat on the choice made. We stress that this is not an unphysical ambiguity, but rather a physical choice of whether to integrate out or not the boundary degree of freedom from the density matrix. We do not expect this to significantly affect the entanglement entropy. In the following we choose to reduce the density matrix over $\phi(x)$ as we want to integrate out {\it all} local gauge invariant degrees of freedom in $R$.  

\section{The reduced density matrix and the entropy}
 In terms of the reduced density matrix $\hat\rho_L$, the von Neumann entropy is defined as

\begin{align}\label{reddens}
S_{E}=-Tr[\rho_Llog(\rho_L)]
\end{align}



Since our decomposition of degrees of freedom preserves translational invariance in the transverse plain, we can conveniently work in the 2d momentum space. 
We use
\begin{equation}
\frac{1}{2\pi}\int d^2(x-y)  \frac{1}{[(x-y)^2+(x_3-y_3)^2]}e^{ik\cdot(x-y)}= K_0[k(x_3-y_3)];
\end{equation}
where $K_0$ is the Bessel K-function and $k$ is the length of the transverse momentum vector.
The transverse and longitudinal fields decouple in the wave function
\beq\label{wf}
\psi=\psi_\chi\psi_\zeta
\eeq
with
\begin{eqnarray}
\psi_\zeta&=&N_\zeta exp\Big\{-\int dzdz'd^2k\ \ \ \ \left[\zeta(k,z)\frac{1}{k^2}K_0(k(z-z'))\zeta(-k,z')\right]\Big\}\\ 
\psi_\chi&=&N_\chi\exp\Big\{-\int dzdz' d^2k\left[\chi(k,z)\frac{1}{k^2}K_0(k(z-z'))\chi(-k,z')+[\phi(k)-\int_0^z du\chi(k,u)]K_0(k(z-z'))[\phi(-k)-\int_0^{z'}dv\chi(-k,v)]\right]\Big\}\nonumber
\end{eqnarray}

We now decompose the fields as
\begin{eqnarray}
\chi(k,z)=\chi_L(k,z)\Theta(-z)+\chi_R(k,z)\Theta(z);\\
\zeta(k,z)=\zeta_L(k,z)\Theta(-z)+\zeta_R(k,z)\Theta(z);\nonumber
\end{eqnarray}
Our next goal is to integrate the density matrix over $\chi_R$, $\zeta_R$ and $\phi$. Since $\chi$ and $\zeta$ decouple, we will consider the two wave functions in turn. Also, since the density matrix is diagonal in transverse momentum space we consider a single transverse momentum mode, and will integrate the entropy over the transverse momentum in the last step of the calculation.

\subsection{The transverse field}
The reduced density matrix for the $\zeta_L$ field is
\begin{eqnarray}\label{rhol}
&&\rho[\zeta_L,\zeta'_L]=\int D\zeta_R\psi_\zeta[\zeta_L,\zeta_R]\psi_\zeta[\zeta'_L,\zeta_R]\nonumber\\
&&=N\int D\zeta_R \exp\Big[-\Big\{\int_{LL}\zeta_L(k,z)\frac{1}{k^2}K_0(k(z-z'))\zeta_L(-k,z')+\int_{LL}\zeta'_L(k,z)\frac{1}{k^2}K_0(k(z-z'))\zeta'_L(-k,z')\\
&&+2\int_{RR}\zeta_R(k,z)\frac{1}{k^2}K_0(k(z-z'))\zeta_R(-k,z')+2\int_{LR}[\zeta_L(k,z)+\zeta'_L(k,z)]\frac{1}{k^2}K_0(k(z-z'))\zeta_R(-k,z')
\Big\}\Big]\nonumber
\end{eqnarray}
Here we have introduced notation $\int_R\equiv\int_0^\infty dz$ and similarly $\int_L\equiv\int_{-\infty}^0dz$.

Since the integral over $\zeta_R$ is Gaussian, the integration is equivalent to solving classical equation of motion for $\zeta_R$
\begin{equation}\label{clas}
\int_{z'>0}K_0(k(z-z'))\zeta_R(k,z')=-\frac{1}{2}\int_{z'<0}K_0(k(z-z'))[\zeta_L(k,z')+\zeta'_L(k,z')],\ \ \ \ \ z>0
\end{equation}
To solve this equation one would need to find an inverse to the modified Bessel function $K_0$ on half a space. Instead of trying to do that exactly, in the following we will  assume that the typical scale of variation of the fields  in the longitudinal direction is greater than $1/k$.  This will allow us to use the asymptotic form of the Bessel function $K_0$ and perform the calculation analytically. Although this is certainly an approximation, we believe that it captures the salient features of  the exact result.

 In this approximation  we can use the asymptotic form of $K_0$ in the right hand side of the equation
\begin{equation}\label{ass}
K_0(x)\rightarrow_{x\gg 1}\sqrt{\frac{\pi}{2x}}e^{-x}
\end{equation}
The integration over $z'$ on the left hand side of eq.(\ref{clas}) is dominated by the region close to the point $z$, and we therefore approximate $K_0$ by a delta function. Given that
\begin{equation}
\int dz K_0(z)=\pi
\end{equation}
we will use on the left hand side of eq.(\ref{clas})
\begin{equation}\label{k0}
K_0(kz)\approx \frac{\pi}{k} \delta(z)
\end{equation}
In this approximation we obtain
\begin{equation}
\frac{\pi}{k}\zeta_R(k,z)=-\frac{1}{2}\int_{z'<0} \sqrt{\frac{\pi}{2k(z-z')}}e^{-k(z-z')}[\zeta_L(k,z')+\zeta'_L(k,z')]\approx-\sqrt{\frac{\pi}{8kz}}e^{-kz}\frac{1}{k}[\zeta_L(k,0)+\zeta'_L(k,0)]
\end{equation}
or
\begin{equation}
\zeta_R(k,z)\approx-\sqrt{\frac{1}{8\pi kz}}e^{-kz}[\zeta_L(k,0)+\zeta'_L(k,0)]
\end{equation}
Using the same local approximation eq.(\ref{k0}) also for the other terms in eq.(\ref{rhol}) we obtain for the density matrix
\begin{eqnarray}\label{rholl}
\rho[\zeta_L,\zeta'_L]&=&N\exp\Big\{-\frac{\pi}{k^3}\Big[\int_{z<0}[\zeta_L(k,z)\zeta_L(-k,z)+\zeta'_L(k,z)\zeta'_L(-k,z)]\\
&-&\int_{z>0}\frac{1}{8\pi kz}e^{-2kz}[\zeta_L(k,0)+\zeta'_L(k,0)][\zeta_L(-k,0)+\zeta'_L(-k,0)]\Big]\Big\}\nonumber\\
&=&N\exp\Big\{-\frac{\pi}{k^3}\Big[\int_{z<0}[\zeta_L(k,z)\zeta_L(-k,z)+\zeta'_L(k,z)\zeta'_L(-k,z)]-\frac{\gamma}{k}[\zeta_L(k,0)+\zeta'_L(k,0)][\zeta_L(-k,0)+\zeta'_L(-k,0)]\Big\}\nonumber
\end{eqnarray}
Here $\gamma$ is a pure number. Our local approximation formally yields a logarithmically divergent value of $\gamma$:
\begin{equation}\label{gamma}
\gamma=\int_{z>0}\frac{dz}{8\pi z}e^{-2z}
\end{equation}
  However this divergence is an artifact of our use of the asymptotic form of the $K$-function eq.(\ref{ass}). The integral in eq.(\ref{gamma}) is dominated by  small values of $z$, and thus the asymptotic form of $K_0$ is not relevant. The actual behavior of $K_0(z)$ for small $z$ is logarithmic, and as a result $\zeta_R(z)$ is finite as $z\rightarrow 0$. A better representation for $\gamma$ is therefore
\begin{equation}
\gamma=\frac{1}{4\pi^2}\int_{z>0}K_0^2(z)
\end{equation}

Note that as long as we consider slowly varying fields $\zeta_L$, the functional form of the last term in eq.(\ref{rholl}) is correct irrespective of the value of $\gamma$. The constant $\gamma$ unfortunately cannot be reliably calculated in the local approximation. We will therefore continue our discussion keeping its value unspecified. 

We can now calculate the entanglement entropy due to the transverse field $\zeta$. To do this in the most efficient way we notice that the only nontrivial contribution to entropy comes from the mode at $z=0$, as for all the other modes our density matrix is that of a pure state. Thus for the purpose of calculating the entropy we can completely disregard all the modes at $z\ne 0$. Of course one has to be careful since longitudinal coordinate is continuous. However all our simplifications so far assumed that the fields only vary on longitudinal scales larger than $1/k$. In other words $k$ serves as a longitudinal UV cutoff on our calculation. We thus proceed taking $1/k$ as the longitudinal "lattice spacing" and discretizing the integral over $z$ accordingly.  The density matrix for the $z=0$ mode then becomes 
\begin{equation}
\rho(\zeta,\zeta')=N\exp\Big\{-\frac{\pi}{k^4}\Big[\zeta_L\zeta^*_L+\zeta'_L\zeta{'^*}_L-\gamma[\zeta_L+\zeta'_L][\zeta^*_L+\zeta{'^*}_L]\Big]\Big\}
\end{equation}
The entropy for this density matrix is readily calculated. Using the results of \cite{Kovner2015a},\cite{Armesto:2019mna} we get
\begin{equation}\label{sz}
S_\zeta=\ln\frac{2\gamma}{4(1-2\gamma)}+\frac{1}{\sqrt{1-2\gamma}}\rm{arcosh}\left[\frac{2-2\gamma}{2\gamma}\right]
\end{equation}
The entropy per transverse momentum mode is obviously finite. Moreover it is independent of transverse momentum $k$. The total entropy is obtained by summing over all modes $k$. To do this one clearly needs to regulate the summation both in the infrared and ultraviolet. Assuming that the transverse momenta are quantized in units of $2\pi/L_\perp$ and attain the maximal value of $2\pi/a$, where $L_\perp$ is the transverse size of the system while $a$ is the ultraviolet cutoff,
we obtain for the entropy due to transverse modes
\begin{equation}\label{tran}
S_T =\frac{L_\perp^2}{a^2}\left[\ln\frac{2\gamma}{4(1-2\gamma)}+\frac{1}{\sqrt{1-2\gamma}}\rm{arcosh}\left[\frac{2-2\gamma}{2\gamma}\right]\right]
\end{equation}
Note that the ultraviolet cutoff $a$ is  determined by the details of separation between the left and right subspaces. Only transverse momenta smaller than the inverse longitudnal size of the boundary can be treated in the approximation we have been using, and thus this longitudinal size plays the role of the ultraviolet cutoff $a$.

The result eq.(\ref{tran}) is quite intuitive. The entropy is extensive and proportional to the number of degrees of freedom on the boundary between the two regions. The entropy per degree of freedom is finite, since the entanglement of transverse degrees of freedom is short range.

\subsection{Longitudinal fields}
We now turn to the evaluation of the entropy of the longitudinal fields.
The density matrix for the longitudinal fields is
\begin{eqnarray}
\rho[\chi_L,\chi'_L]&=&N_\chi\int D\chi_RD\phi\exp\Bigg\{-\int_{LL}\Bigg[\chi_L(k,z)\left[\frac{1}{k^2}K_0(k(z-z'))+\int_{u<z,v<z'}K_0(k(u-v))\right]\chi_L(-k,z')\nonumber\\
&+&\chi'_L(k,z)\left[\frac{1}{k^2}K_0(k(z-z'))+\int_{u<z,v<z'}K_0(k(u-v))\right]\chi'_L(-k,z')\Bigg]\nonumber\\
&&-2\int_{RR}\chi_R(k,z)\left[\frac{1}{k^2}K_0(k(z-z'))+\int_{u>z,v>z'}K_0(k(u-v))\right]\chi_R(-k,z')\\
&&-2\int_{LR}[\chi_L(k,z)+\chi'_L(k,z)]\left[\frac{1}{k^2}K_0(k(z-z'))-\int_{u<z,v>z'}K_0(k(u-v))\right]\chi_R(-k,z')\nonumber\\
&&-\frac{4\pi L}{k}\phi(k)\phi(-k)+\frac{4\pi}{k}\phi(k)\int_{R}[L-u]\chi_R(-k,u)-\frac{2\pi}{k}\phi(k)\int_{L}[L+u][\chi_L(-k,u)+\chi'_L(-k,u)]\Bigg\}\nonumber
\end{eqnarray}

where $L$ is the infrared cutoff on the longitudinal size of the system. Here we dropped the integral over the transverse momentum $k$, as different $k$-modes decouple and we calculate the entropy for each mode separately.

As before, we now approximate the $K_0$ function by a delta function whenever it appears between two fields which are both either in the left or right part of space.  Our expression then becomes
\begin{eqnarray}
\rho[\chi_L,\chi'_L]&=&N_\chi\int D\chi_RD\phi\exp\Bigg\{-\frac{\pi}{k^3}\int_L\chi_L(k,z)\chi_L(-k,z)-\frac{2\pi}{k}\int_z\chi_L(k,z)[L+z]\int_{z'>z}\chi_L(-k,z')\nonumber\\
&-&\frac{\pi}{k^3}\int_L\chi'_L(k,z)\chi'_L(-k,z)-\frac{2\pi}{k}\int_z\chi'_L(k,z)[L+z]\int_{z'>z}\chi'_L(-k,z')\\
&&-\frac{2\pi}{k^3}\int_{R}\chi_R(k,z)\chi_R(-k,z)-\frac{4\pi}{k}\int_R \chi_R(k,z)[L-z]\int_{z'<z}\chi_R(-k,z')\nonumber\\
&&-2\int_{LR}[\chi_L(k,z)+\chi'_L(k,z)]\left[\frac{1}{k^2}K_0(k(z-z'))-\int_{u<z,v>z'} K_0(k(u-v))\right]\chi_R(-k,z')\nonumber\\
&&-\frac{4\pi L}{k}\phi(k)\phi(-k)+\frac{4\pi}{k}\phi(k)\int_{R}[L-u]\chi_R(-k,u)-\frac{2\pi}{k}\phi(k)\int_{L}[L+u][\chi_L(-k,u)+\chi'_L(-k,u)]\Bigg\}\nonumber
\end{eqnarray}
The classical equations for $\phi$ and $\chi_R$ are
\begin{eqnarray}\label{eqas}
2L\phi&=&\int_R(L-u)\chi_R-\frac{1}{2}\int_L(L+u)[\chi_L+\chi'_L]\\
\frac{1}{k^2}\chi_R(z)&+&[L-z]\int_{z'<z}\chi_R(z')+\int_{z'>z}[L-z']\chi_R(z')-[L-z]\phi\nonumber\\
&+&\frac{1}{2\pi k}\int_LK_0(k(z-z')[\chi_L(z')+\chi'_L(z')]-\frac{k}{2\pi}\int_{v>z; \ u<z';\ z'}K_0(k(v-u))[\chi_L(z')+\chi'_L(z')]=0\nonumber
\end{eqnarray}
We note that for the asymptotic form of the $K_0$ function we have
\begin{equation}
\int_{v>z; \ u<z'}K_0(k(v-u))=\frac{1}{k^2}K_0(k(z-z'))
\end{equation}
This simplifies our equations considerably and has the effect of cancelling $\chi_L$ dependent terms in the second equation in eq.(\ref{eqas}). Our equations now become 
\begin{eqnarray}\label{equa}
2L\phi&=&\int_R(L-u)\chi_R-\frac{1}{2}\int_L(L+u)[\chi_L+\chi'_L]\nonumber\\
\frac{1}{k^2}\chi_R(z)&+&[L-z]\int_{z'<z}\chi_R(z')+\int_{z'>z}[L-z']\chi_R(z')-[L-z]\phi=0
\end{eqnarray}

This can be easily solved. Differentiating eq.(\ref{equa})  twice with respect to $z$ we obtain.
\begin{equation}
\frac{d^2}{dz^2}\chi_R-k^2\chi_R(z)=0
\end{equation}
Thus
\begin{equation}
\chi_R(z)=\chi_R(0)e^{-kz}
\end{equation}
To determine the constant we substitute this solution back into eq.(\ref{equa})  with the result
\begin{equation}
\chi_R(0)=-\frac{k^2}{2}\frac{1}{Lk+1}\int_L[L+u][\chi_L(u)+\chi'_L(u)]
\end{equation}
so that finally the solution of classical equations is
\begin{eqnarray}
&&\chi_R(z)=-\frac{k^2}{2}\frac{1}{Lk+1}e^{-kz}\int_L[L+u][\chi_L(u)+\chi'_L(u)]\\
&&\phi=-\frac{k}{2}\frac{1}{Lk+1}\int_L[L+u][\chi_L(u)+\chi'_L(u)]\nonumber
\end{eqnarray}
The reduced density matrix now is 
\begin{eqnarray}
\rho[\chi_L,\chi'_L]&=&N_\chi\exp\Bigg\{-\frac{\pi}{k^3}\int_L\chi_L(k,z)\chi_L(-k,z)-\frac{2\pi}{k}\int_z\chi_L(k,z)[L+z]\int_{z'>z}\chi_L(-k,z')\nonumber\\
&-&\frac{\pi}{k^3}\int_L\chi'_L(k,z)\chi'_L(-k,z)-\frac{2\pi}{k}\int_z\chi'_L(k,z)[L+z]\int_{z'>z}\chi'_L(-k,z')\nonumber\\
&+&\frac{\pi}{2}\frac{Lk+\frac{1}{2}}{(Lk+1)^2}\int_L[L+u][\chi_L(k,u)+\chi'_L(k,u)]\int_L[L+u][\chi_L(-k,u)+\chi'_L(-k,u)]\Bigg\}
\end{eqnarray}
If we further assume that the longitudinal fields vanish at infinity, so that the longitudinal integrals are finite, we can neglect the factors of $z$ whenever they appear in the sum $L+z$. We then get
\begin{eqnarray}
\rho[\chi_L,\chi'_L]&\approx&N_\chi\exp\Bigg\{-\frac{\pi}{k^3}\int_L\chi_L(k,z)\chi_L(-k,z)-\frac{2\pi L}{k}\int_L\chi_L(k,z)\int_L\chi_L(-k,z')\nonumber\\
&-&\frac{\pi}{k^3}\int_L\chi'_L(k,z)\chi'_L(-k,z)-\frac{2\pi L}{k}\int_z\chi'_L(k,z)\int_L\chi'_L(-k,z')\nonumber\\
&+&\frac{\pi L}{2k}\int_L[\chi_L(k,u)+\chi'_L(k,u)]\int_L[\chi_L(-k,u)+\chi'_L(-k,u)]\Bigg\}
\end{eqnarray}
Interestingly, just like in the case of the transverse fields, for longitudinal fields it is only one mode that dominates the entanglement properties between the right and left spaces (at each value of $k$). The  difference is that now the mode in question is not a local mode at the boundary, but rather the integral of the magnetic field over the half space. This difference comes about due the necessity of solving the no monopole constraint for the longitudinal magnetic field.

To calculate the entropy we now represent the longitudinal field as
\begin{equation}\label{chi0}
\chi_L(z)=\tilde\chi_L(z)+\chi^0_L; \ \ \ \chi^0_L=\frac{1}{L}\int_L\chi_L(z);\ \ \ \ \ \int_L\tilde\chi_L(z)=0
\end{equation}
It is readily seen that $\chi^0$ and $\tilde\chi$ decouple in the density matrix and the only entangled factor in $\rho$ is associated with $\chi^0_L$:
\begin{eqnarray}
\rho_0[\chi^0,\chi^{0'}]&\approx &N_0\exp\{-\frac{\pi L}{k^3}\left[\left(1+2L^2k^2\right)\left[\chi^0_L(k)\chi^0_L(-k)+\chi^{0'}_L(k)\chi^{0'}_L(-k)\right]-\frac{L^2k^2}{2}[\chi^0_L(k)+\chi^{0'}_L(k)][\chi^0_L(-k)+\chi^{0'}_L(-k)]\right]\}\nonumber\\
&\approx&N_0\exp\{-\frac{\pi L^3}{2k}\left[[\chi^0_L(k)+\chi^{0'}_L(k)][\chi^0_L(-k)+\chi^{0'}_L(-k)]+2[\chi^0_L(k)-\chi^{0'}(k)][\chi^0_L(-k)-\chi^{0'}(-k)]\right]\}
\end{eqnarray}
Finally again using \cite{Armesto:2019mna}, we get for the entanglement entropy:
\begin{equation}
S_\chi=\left[\ln\frac{1}{4}+\sqrt{2}\ {\rm{ arcosh}} \ 3\right]
\end{equation}

Summing over the transverse momentum modes we obtain
\begin{equation}\label{long}
S_\chi=\frac{L_\perp^2}{a^2}\left[\ln\frac{1}{4}+\sqrt{2}\ {\rm{ arcosh}} \ 3\right]
\end{equation}

\section{Discussion}
We now summarize our results. We find that the entanglement entropy is
proportional to the transverse area of the plane separating the two halves of space.
This as it should be, since entropy is an extensive quantity. In fact the factor $L_\perp^2/a^2$ has a natural interpretation as the number of degrees of freedom on the boundary between the two regions. 

For the contribution of transverse modes one indeed can straightforwardly interpret the result eq.(\ref{tran}) in this way with eq.(\ref{sz}) being the entropy per degree of freedom on the boundary. The entanglement between left and right degrees of freedom here is short range and is entirely localized in the vicinity of the boundary. 

The contribution of the longitudinal modes is of a somewhat different nature.
Their entanglement  as we have seen is due to the constant mode in the longitudinal direction  and thus cannot be attributed  to the boundary region. Examining  our calculation closely it is clear that this long range entanglement originates from  the  solution of the no monopole condition for the magnetic field. The integral expressing the third component of magnetic field $B_3$ in terms of $B_i$
propagates all the way through our calculations and is the reason why the entanglement is concentrated in the nonlocal mode $\chi^0$ defined in eq.(\ref{chi0}). 
In fact one has $\chi^0(x)\propto B_3(\infty, x)-B_3(0,x)$. It is thus natural to attribute the long range entanglement to the $z$ component of magnetic field, although one has to keep in mind that such an identification is basis dependent.

We note here that since the entropy is contributed by regions far away from the boundary, we believe that the local approximation we employed throughout this paper is robust and correctly reflects the nature of entanglement.


It would be very interesting to see whether the long range entanglement is also present in non Abelian theories, such as pure gluodynamics. On one hand these theories do contain a differential condition constraining the magnetic field, albeit this condition involves a covariant divergence. On the other hand non Abelian theories have a finite mass gap or equivalently,  finite correlation length. It would be  very surprising if spatial regions separated in the longitudinal direction by a distance greater than the correlation length would be entangled in the vacuum wave function. We therefore may surmise that in non Abelian theories the entanglement in the longitudinal direction is cut off on distances of order of the correlation length. This effect must be nonperturbative and therefore difficult to uncover.

\acknowledgements
We thank Mahesh Chandran for pointing out a mistake in the first version of this paper.

\bibliography{bibliography}

\begin{thebibliography}{16}
\expandafter\ifx\csname natexlab\endcsname\relax\def\natexlab#1{#1}\fi
\expandafter\ifx\csname bibnamefont\endcsname\relax
  \def\bibnamefont#1{#1}\fi
\expandafter\ifx\csname bibfnamefont\endcsname\relax
  \def\bibfnamefont#1{#1}\fi
\expandafter\ifx\csname citenamefont\endcsname\relax
  \def\citenamefont#1{#1}\fi
\expandafter\ifx\csname url\endcsname\relax
  \def\url#1{\texttt{#1}}\fi
\expandafter\ifx\csname urlprefix\endcsname\relax\def\urlprefix{URL }\fi
\providecommand{\bibinfo}[2]{#2}
\providecommand{\eprint}[2][]{\url{#2}}

\bibitem[{\citenamefont{Kitaev and Preskill}(2006)}]{Kitaev:2005dm}
\bibinfo{author}{\bibfnamefont{A.}~\bibnamefont{Kitaev}} \bibnamefont{and}
  \bibinfo{author}{\bibfnamefont{J.}~\bibnamefont{Preskill}},
  \bibinfo{journal}{Phys. Rev. Lett.} \textbf{\bibinfo{volume}{96}},
  \bibinfo{pages}{110404} (\bibinfo{year}{2006}), \eprint{hep-th/0510092}.

\bibitem[{\citenamefont{Levin and Wen}(2006)}]{PhysRevLett.96.110405}
\bibinfo{author}{\bibfnamefont{M.}~\bibnamefont{Levin}} \bibnamefont{and}
  \bibinfo{author}{\bibfnamefont{X.-G.} \bibnamefont{Wen}},
  \bibinfo{journal}{Phys. Rev. Lett.} \textbf{\bibinfo{volume}{96}},
  \bibinfo{pages}{110405} (\bibinfo{year}{2006}),
  \urlprefix\url{https://link.aps.org/doi/10.1103/PhysRevLett.96.110405}.

\bibitem[{\citenamefont{Witten}(2018)}]{Witten:2018lha}
\bibinfo{author}{\bibfnamefont{E.}~\bibnamefont{Witten}},
  \bibinfo{journal}{Rev. Mod. Phys.} \textbf{\bibinfo{volume}{90}},
  \bibinfo{pages}{045003} (\bibinfo{year}{2018}), \eprint{1803.04993}.

\bibitem[{\citenamefont{Solodukhin}(2008)}]{Solodukhin:2008dh}
\bibinfo{author}{\bibfnamefont{S.~N.} \bibnamefont{Solodukhin}},
  \bibinfo{journal}{Phys. Lett.} \textbf{\bibinfo{volume}{B665}},
  \bibinfo{pages}{305} (\bibinfo{year}{2008}), \eprint{0802.3117}.

\bibitem[{\citenamefont{Casini and Huerta}(2009)}]{Casini:2009sr}
\bibinfo{author}{\bibfnamefont{H.}~\bibnamefont{Casini}} \bibnamefont{and}
  \bibinfo{author}{\bibfnamefont{M.}~\bibnamefont{Huerta}},
  \bibinfo{journal}{J. Phys.} \textbf{\bibinfo{volume}{A42}},
  \bibinfo{pages}{504007} (\bibinfo{year}{2009}), \eprint{0905.2562}.

\bibitem[{\citenamefont{Casini et~al.}(2014)\citenamefont{Casini, Huerta, and
  Rosabal}}]{Casini:2013rba}
\bibinfo{author}{\bibfnamefont{H.}~\bibnamefont{Casini}},
  \bibinfo{author}{\bibfnamefont{M.}~\bibnamefont{Huerta}}, \bibnamefont{and}
  \bibinfo{author}{\bibfnamefont{J.~A.} \bibnamefont{Rosabal}},
  \bibinfo{journal}{Phys. Rev.} \textbf{\bibinfo{volume}{D89}},
  \bibinfo{pages}{085012} (\bibinfo{year}{2014}), \eprint{1312.1183}.

\bibitem[{\citenamefont{Casini and Huerta}(2014)}]{Casini:2014aia}
\bibinfo{author}{\bibfnamefont{H.}~\bibnamefont{Casini}} \bibnamefont{and}
  \bibinfo{author}{\bibfnamefont{M.}~\bibnamefont{Huerta}},
  \bibinfo{journal}{Phys. Rev.} \textbf{\bibinfo{volume}{D90}},
  \bibinfo{pages}{105013} (\bibinfo{year}{2014}), \eprint{1406.2991}.

\bibitem[{\citenamefont{Huang}(2015)}]{Huang:2014pfa}
\bibinfo{author}{\bibfnamefont{K.-W.} \bibnamefont{Huang}},
  \bibinfo{journal}{Phys. Rev.} \textbf{\bibinfo{volume}{D92}},
  \bibinfo{pages}{025010} (\bibinfo{year}{2015}), \eprint{1412.2730}.

\bibitem[{\citenamefont{Huerta and Pedraza}(2018)}]{Huerta:2018xvl}
\bibinfo{author}{\bibfnamefont{M.}~\bibnamefont{Huerta}} \bibnamefont{and}
  \bibinfo{author}{\bibfnamefont{L.~A.} \bibnamefont{Pedraza}}
  (\bibinfo{year}{2018}), \eprint{1808.01864}.

\bibitem[{\citenamefont{Kabat}(1995)}]{Kabat:1995eq}
\bibinfo{author}{\bibfnamefont{D.~N.} \bibnamefont{Kabat}},
  \bibinfo{journal}{Nucl. Phys.} \textbf{\bibinfo{volume}{B453}},
  \bibinfo{pages}{281} (\bibinfo{year}{1995}), \eprint{hep-th/9503016}.

\bibitem[{\citenamefont{Donnelly and Wall}(2016)}]{Donnelly:2015hxa}
\bibinfo{author}{\bibfnamefont{W.}~\bibnamefont{Donnelly}} \bibnamefont{and}
  \bibinfo{author}{\bibfnamefont{A.~C.} \bibnamefont{Wall}},
  \bibinfo{journal}{Phys. Rev.} \textbf{\bibinfo{volume}{D94}},
  \bibinfo{pages}{104053} (\bibinfo{year}{2016}), \eprint{1506.05792}.

\bibitem[{\citenamefont{Agarwal et~al.}(2017)\citenamefont{Agarwal, Karabali,
  and Nair}}]{Agarwal:2016cir}
\bibinfo{author}{\bibfnamefont{A.}~\bibnamefont{Agarwal}},
  \bibinfo{author}{\bibfnamefont{D.}~\bibnamefont{Karabali}}, \bibnamefont{and}
  \bibinfo{author}{\bibfnamefont{V.~P.} \bibnamefont{Nair}},
  \bibinfo{journal}{Phys. Rev.} \textbf{\bibinfo{volume}{D96}},
  \bibinfo{pages}{125008} (\bibinfo{year}{2017}), \eprint{1701.00014}.

\bibitem[{\citenamefont{Donnelly}(2012)}]{Donnelly:2011hn}
\bibinfo{author}{\bibfnamefont{W.}~\bibnamefont{Donnelly}},
  \bibinfo{journal}{Phys. Rev.} \textbf{\bibinfo{volume}{D85}},
  \bibinfo{pages}{085004} (\bibinfo{year}{2012}), \eprint{1109.0036}.

\bibitem[{\citenamefont{Radičević}(2016)}]{Radicevic:2015sza}
\bibinfo{author}{\bibfnamefont{Ã.}~\bibnamefont{Radičević}},
  \bibinfo{journal}{JHEP} \textbf{\bibinfo{volume}{04}}, \bibinfo{pages}{163}
  (\bibinfo{year}{2016}), \eprint{1509.08478}.

\bibitem[{\citenamefont{Kovner and Lublinsky}(2015)}]{Kovner2015a}
\bibinfo{author}{\bibfnamefont{A.}~\bibnamefont{Kovner}} \bibnamefont{and}
  \bibinfo{author}{\bibfnamefont{M.}~\bibnamefont{Lublinsky}},
  \bibinfo{journal}{Physical Review D - Particles, Fields, Gravitation and
  Cosmology} \textbf{\bibinfo{volume}{92}}, \bibinfo{pages}{1}
  (\bibinfo{year}{2015}), ISSN \bibinfo{issn}{15502368},
  \eprint{arXiv:1506.05394v1}.

\bibitem[{\citenamefont{Armesto et~al.}(2019)\citenamefont{Armesto, Dominguez,
  Kovner, Lublinsky, and Skokov}}]{Armesto:2019mna}
\bibinfo{author}{\bibfnamefont{N.}~\bibnamefont{Armesto}},
  \bibinfo{author}{\bibfnamefont{F.}~\bibnamefont{Dominguez}},
  \bibinfo{author}{\bibfnamefont{A.}~\bibnamefont{Kovner}},
  \bibinfo{author}{\bibfnamefont{M.}~\bibnamefont{Lublinsky}},
  \bibnamefont{and} \bibinfo{author}{\bibfnamefont{V.}~\bibnamefont{Skokov}}
  (\bibinfo{year}{2019}), \eprint{1901.08080}.

\end{thebibliography}

\end{document}